# A 20 Gbps PAM4 Data Transmitter ASIC for Particle Physics Experiments

**L. Zhang,**[a,b,*] **E.M. Cruda,**[a] **C-P. Chao,**[c] **S-W. Chen,**[c] **B. Deng,**[d] **R. Francisco,**[f] **D.Gong,**[a,†] **D. Guo,**[a] **S. Hou,**[e] **G. Huang,**[b] **X. Huang,**[a] **S. Kulis,**[f] **C-Y. Li,**[c] **C. Liu,**[a] **E.R. Liu,**[i] **T. Liu,**[a] **P. Moreira,**[f] **J. Prinzie,**[h] **H. Sun,**[a,b,*] **Q. Sun,**[g] **X. Sun,**[b] **G. Wong,**[j] **D. Yang,**[a] **J. Ye,**[a] **and W. Zhang,**[a,b,*]

[a] *Southern Methodist University, Dallas, TX 75275, U.S.A.*
[b] *Central China Normal University, Wuhan, Hubei 430079, P.R. China*
[c] *APAC, Opto Electronics Inc., Hukow, Hsinchu 303, Taiwan*
[d] *Hubei Polytechnic University, Huangshi, Hubei 435003, P.R. China*
[e] *Academia Sinica, Nangang, Taipei 11529, Taiwan*
[f] *CERN, 1211 Geneva 23, Switzerland*
[g] *Fermi National Accelerator Laboratory, Batavia, IL 60510, U.S.A.*
[h] *ESAT-ADVISE research lab, KU Leuven University, 3000 Leuven, Belgium*
[i] *Allen High School, Allen, TX 75002, U.S.A.*
[j] *Plano East Senior High School, Plano, TX 75074, U.S.A.*

*E-mail*: dgong@smu.edu

ABSTRACT: We present the design and test results of a novel data transmitter ASIC operating up to 20.48 Gbps with 4-level Pulse-Amplitude-Modulation (PAM4) for particle physics experiments. This ASIC, named GBS20, is fabricated in a 65 nm CMOS technology. Two serializers share a 5.12 GHz Phase Locked Loop (PLL) clock. The outputs from the serializers are combined into a PAM4 signal that directly drives a Vertical-Cavity-Surface-Emitting-Laser (VCSEL). The input data channels, each at 1.28 Gbps, are scrambled with an internal $2^7-1$ Pseudo-Random Binary Sequence (PRBS), which also serves as a frame aligner. GBS20 is tested to work at 10.24 and 20.48 Gbps with a VCSEL-based Transmitter-Optical-Subassembly (TOSA). The power consumption of GBS20 is below 238 mW and reduced to 164 mW in the low-power mode.

---

[*] Visiting scholar at SMU and performed this work at SMU

[†] Corresponding authors.

## Contents

## 1. Introduction

The physics experiments being prepared for the coming High-Luminosity Large-Hadron Collider (HL-LHC) demand a higher detector data transmission rate. Currently, the lpGBT [1] ASIC developed for the HL-LHC reaches a 10.24 Gbps non-return-to-zero (NRZ) data rate. We reuse several design blocks from lpGBT and follow the dual-serializer concept that is first used in LOCx2 [2]. A 4-level Pulse-Amplitude-Modulation (PAM4) circuit and a Vertical-Cavity-Surface-Emitting-Laser (VCSEL) driver are integrated into a single ASIC named GBS20. It combines two 10.24 Gbps bitstreams into one 20.48 Gbps PAM4 datastream. With the GBS20, we further propose to develop the optical data transmitter GBT20 and receiver GBR20 in the form of pluggable mezzanine cards so that the high-speed optical link is transparent to users. The overall optical link architecture based on GBT20 and GBR20 is shown in Figure 1. It will greatly simplify the link implementation, reduce the number of fibers (compared with a 10.24 Gbps NRZ transmission) from on-detector to off-detector electronics, and better use the input bandwidth of the FPGA devices.

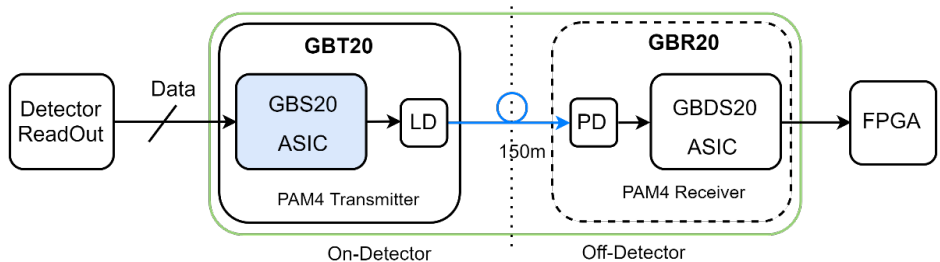

## 2. Design of GBS20

The architecture of GBS20 is shown in Figure 2. Sixteen differential input user-data channels, each at 1.28 Gbps, are received by the electrical receiver (eRx). The Aligner, employing phase-shift and edge-detecting circuits, ensures that the input data is sampled correctly. The input data are divided into two groups, the Least Significant Bits (LSBs) and the Most Significant Bits (MSBs). Each data group is scrambled with an internal $2^7-1$ Pseudo-Random Binary Sequence (PRBS), which also generates the test patterns or frame aligners. The scrambled data is fed to the 8:1 serializer that produces a 5.12 Gbp/s or 10.24 Gbp/s serial bitstream. The Phase-Locked Loop (PLL), which provides low-jitter clocks for all the digital components, including the two serializers and encoders, is derived from lpGBT. This design guarantees a proper phase relationship between the two serial bitstreams. It also removes the need for a Clock-Data Recovery (CDR) circuit that one finds in a standalone PAM4 circuit. The outputs of the two serializers are combined into a PAM4 signal after the five-stage Limiting Amplifier (LA). The latencies of the two groups of LAs are designed to be the same, ensuring that the LSB and MSB have the same phase when they enter into the PAM4 Combiner. The whole chip operates at 1.2 V except the Combiner, which can operate at 1.2 V in low power mode, or 2.5 V to reach a higher modulation current. AC coupling is used to eliminate common-mode voltage mismatch between the LA and Combiner. An Inter-Integrated Circuit ($I^2C$) block is adopted to configure GBS20. Other than the PAM4 Combiner and the Encoder circuits, the rest blocks are from lpGBT and some with adaptations. The power supplies of GBS20 are separated into 1.2 V in the digital domain, 1.2 V and 2.5 V in the analog domain to reduce power noise.

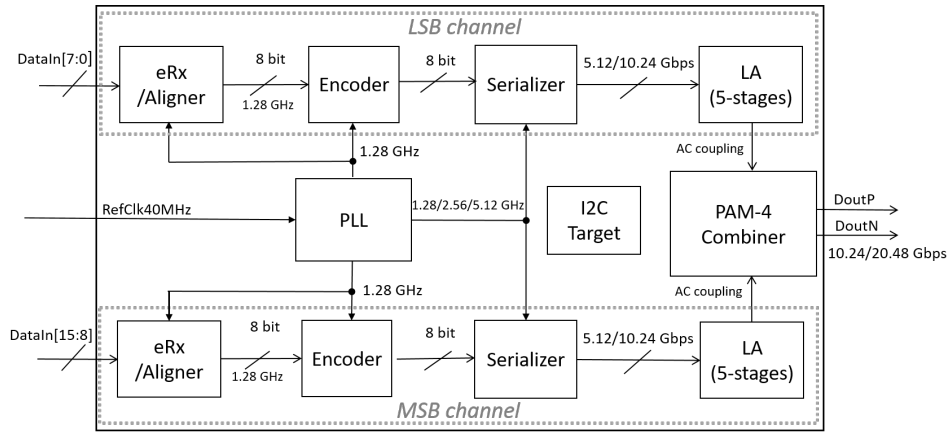

Figure 2. Block diagram of GBS20.

The PAM4 Combiner schematic [3], shown in Figure 3(a), has two fully differential amplifiers sharing a pair of 50 Ω resistors. The tail current in the MSB branch (transistors M8 and M9) is nominally a factor of 2 in the LSB branch (transistors M1 and M2) so that the LSB and MSB are combined to a PAM4 signal. These tail currents are individually adjustable to cope with the nonlinearities of the circuit and the VCSEL. The LA in the LSB or MSB branch can be turned off, providing a way to check the other serializer's NRZ output. The Combiner is designed to run at 1.2 V or 2.5 V. The latter will result in a larger driving current (up to 12 mA) to the VCSEL, producing a more open eye diagram when needed. Passive inductor (L1) peaking and a continuous-time-linear-equalizer (CTLE) with a 3-bit programmable resistor are used to boost the bandwidth. It is difficult to precisely simulate the peaking strength of the passive inductor, and PAM4 signals are much more sensitive to over-peaking than NRZ signals. Therefore, a 5-bit programmable capacitive load is used to fine-tune the overall peaking strength. The LSB and MSB branch have the same structure and similar bandwidths that are above 6.3 GHz. Figure 3(b) shows the post-layout AC simulation result of the LSB side of the Combiner with different CTLE strengths. The maximum peaking gain is about 2.3 dB.

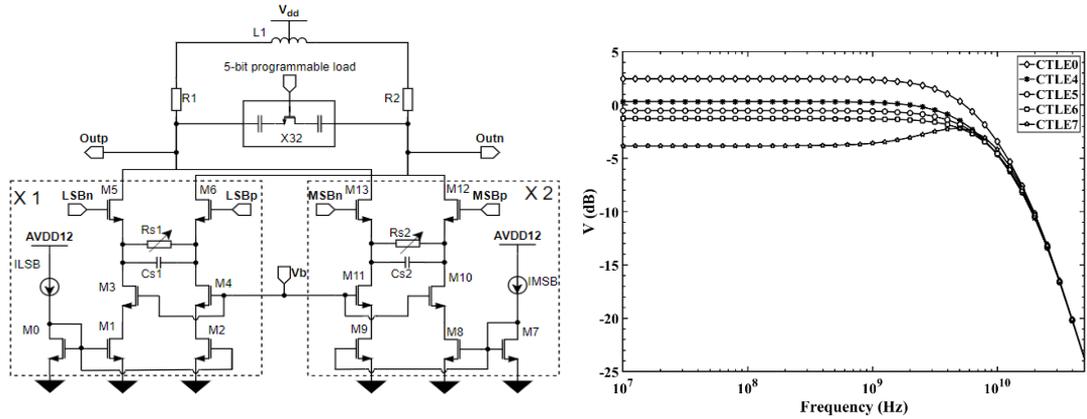

Figure 3. Schematic of the PAM4 combiner (a) and simulated amplitude-frequency response curves with different CTLE strengths (b).

## 3. Test results

The prototype GBS20 is 2 mm × 2 mm and fabricated in a 65 nm CMOS technology. Test boards are assembled with electrical output through SMA connectors or a TOSA to fully characterize all the parameters. A chip wire-bonded to the PCB is shown in Figure 4(a). In Figure 4(b), a 25 Gbps TOSA (Part No.4020180200 from SAN-U) is soldered to the PCB with the GBS20 die under a plastic cover for protection.

A photo of the test setup with a TOSA is in Figure 4(c). A clock board (Silicon Labs Si5338) generates three pairs of 40 MHz synchronized clock signals. One clock is the input for GBS20. The other clock is for a Cyclone 10 GX FPGA development kit (DK-DEV-10CX220-A), which generates the 16 channels of the differential input data to GBS20. The last clock triggers the sampling optical oscilloscope (TDS 8000B with module 80C08C), which measures the TOSA optical output. In addition, a pair of 1.28 GHz test clocks can be measured to evaluate the PLL performance. An I$^2$C level translator board employs a commercial I$^2$C controller to configure GBS20.

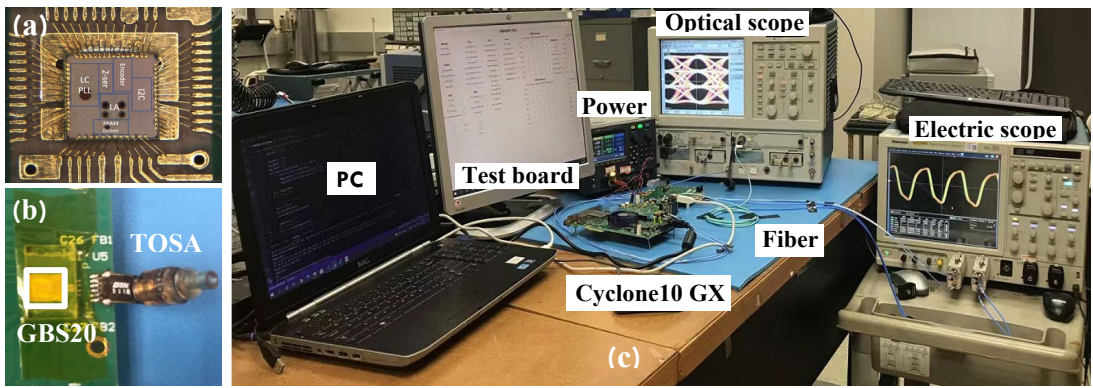

Figure 4. Photos of the GBS20 wire-bonded to the PCB (a) TOSA connected to GBS20 (b) and test setup (c).

The Aligner receives a 1.28 GHz clock from the PLL. Due to a 16-stage voltage-controlled delay line (VCDL) and a 16:1 multiplexer in the phase shift, an internal clock with 48.8 ps phase resolution is created. The aligner circuit function is examined by scanning the generated clock with respect to the input data and checking the output signal's bit error rate (BER). The Aligner scans the data transition with a programmable phase clock and determines the optimal phase of the clock to latch the user data. When the 1.28 Gbps PRBS data generated by the FPGA is aligned with the Aligner's internal clock, the BER is lower than $1 \times 10^{-12}$.

We have checked the LSB and MSB branch independently. Figure 5 shows the optical eye diagrams of LSB (a) and MSB (b) at 10.24 Gbps, with proper adjustment of the peaking strength. As we do not have a PAM4 receiver at this moment, the NRZ outputs also enable us to measure this ASIC's parameters and conduct BER tests for the full transmission chain. This is especially important for the upcoming irradiation tests with single event upset (SEU). The NRZ basic parameters are listed in Table 1.

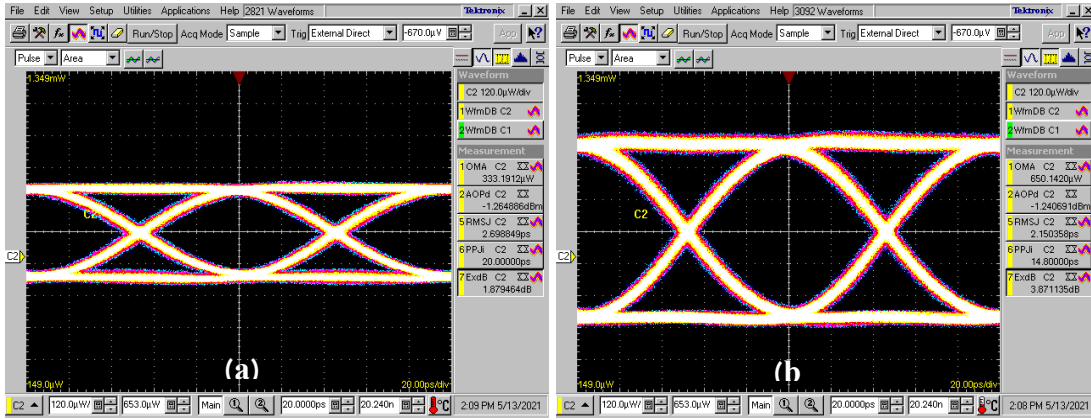

Figure 5. Optical eye diagrams of the LSB channel (a) and MSB channel (b).

The final function tests are performed with the optical eye diagrams at PAM4 through the TOSA. The optical eye diagrams, shown in Figure 6, are measured with the sampling optical oscilloscope that does not have the capability to characterize a PAM4 signal. Through manual measurements, we obtain the parameters in Table 2. A more thorough and industry-standard test with parameters like Transmitter Dispersion Eye Closure Quaternary (TDECQ) will be carried out after we improve our test capabilities.

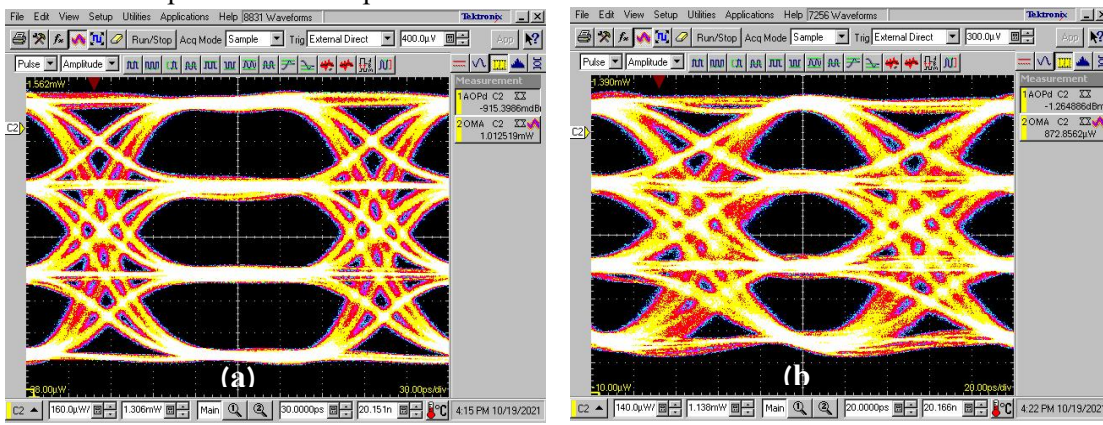

Figure 6. PAM4 optical eyes at 10.24 Gbps (a) and 20.48 Gbps (b).

In the tables AOP stands for average optical power, OMA for optical modulation amplitude, ER for extinction ratio, RT and FT are rise time and fall time, respectively. $OMA_{outer}$ is the envelope OMA and RLM [4] is the transmitter linearity of the PAM4 signal.

Table 1. NRZ optical eye diagram parameter

| CH | AOP (dBm) | OMA (dBm) | ER (dB) | RT (ps) | FT (ps) |
|---|---|---|---|---|---|
| LSB | -1.3 | -4.8 | 1.9 | 58 | 43 |
| MSB | -1.3 | -1.9 | 3.9 | 52 | 42 |

Table 2. PAM4 optical eye diagram parameter

| Data Rate | AOP (dBm) | $OMA_{outer}$ (dBm) | ER (dB) | RLM |
|---|---|---|---|---|
| 10.24G | -0.9 | 1.14 | 9.5 | 0.90 |
| 20.48G | -1.3 | 0.07 | 7.3 | 0.98 |

The power dissipation is less than 238 mW when the PAM4 Combiner works at a supply voltage of 2.5 V. The PAM4 combiner can also operate at 1.2 V (low-power mode). The low-power mode is intended to reduce the power consumption but not significantly compromise performance, especially in radiation tolerance. In the lower-power mode, the power dissipation is reduced to 164 mW. The distribution of the power consumptions in these two modes is shown in Table 3. In our design, all transistors are much larger than the minimum size for radiation tolerance. A Triple Modular Redundancy (TMR) technique is applied in the clock divider and other digital parts to mitigate single event effects. The PAM4 Combiner operates in the current steering mode to drive a VCSEL directly. Therefore, the total power consumptions is higher than similar circuits in industrial applications.

Table 3. Distribution of the power consumption

| Mode | LA+Combiner | PLL | eRx+Encoder+Serialier+$I^2C$ | TOSA's bias | Total |
|---|---|---|---|---|---|
| High-power (mW) | 112 | 63 | 37 | 26 | 238 |
| Low-power (mW) | 51 | 63 | 34 | 16 | 164 |

## 4. Design of GBT20

We currently have three GBT20 mezzanine boards, as shown in Figure 7. The VCSEL is either in a TOSA or under an LC lens. A custom latch (3D printing prototyped) attaches a fiber with a standard LC connector to the VCSEL. Two electrical connectors, the FireFly and the OSFP, are used. The PCB design will be submitted soon.

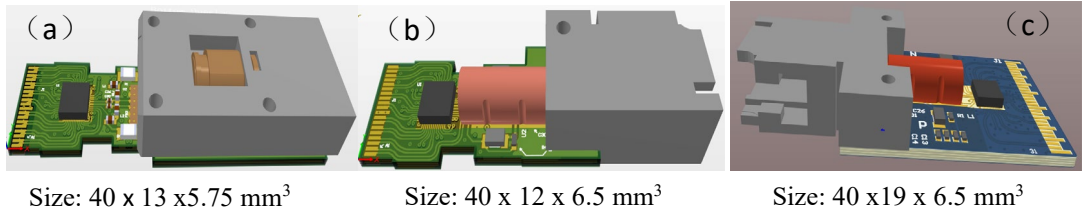

(a) Size: 40 x 13 x5.75 $mm^3$    (b) Size: 40 x 12 x 6.5 $mm^3$    (c) Size: 40 x19 x 6.5 $mm^3$

Figure 7. GBT20 3D models: TOSA with the FireFly connector (a) LC lens with FireFly (b), and LC lens with the OSFP connector (c).

## 5. Conclusion

We present the design and test results of GBS20, our first PAM4 serializer/transmitter for particle physics experiments. Test results show GBS20 passes all the function tests and operates at 20.48 or 10.24 Gbps. More tests, especially irradiation tests, will be carried out in due time. We are also working on the designs of transmitter module GBT20, the receiver ASIC GBDS20, and the receiver module GBR20.

**Acknowledgments**

This work is supported by SMU's Dedman Dean's Research Council Grant and the National Science Council in Taiwan.